\documentclass[showpacs,preprintnumbers]{revtex4}
\usepackage{amssymb}
\usepackage{graphicx}
\usepackage{color}

\def\bc{\begin{center}}
\def\ec{\end{center}}

\def\beq{\begin{equation}}
\def\eeq{\end{equation}}

\input{tcilatex}
\begin{document}

\date{\today}
\title{Charge-exchange, ionization and excitation in low-energy Li$^{+}-$
Ar, \ \ \ K$^{+}-$ Ar, and Na$^{+}-$He collisions.}
\author{ Ramaz A. Lomsadze$^{1}$}
\author{Malkhaz R. Gochitashvili$^{1}$}
\author{RomanYa. Kezerashvili$^{2,3}$ }
\email{rkezerashvili@citytech.cuny.edu }
\author{Michael Schulz$^{4}$ }
\affiliation{$^{1}$ Tbilisi State University, Tbilisi, 0128, Georgia}
\affiliation{$^{2}$ Physics Department, New York City College of Technology, The City
University of New York, Brooklyn, NY 11201, USA}
\affiliation{$^{3}$ The Graduate School and University Center, The City University of New
York, New York, NY 10016, USA}
\affiliation{$^{4}$ Missouri University of Science and Technology, Rolla, MO 65409, USA}

\begin{abstract}
Absolute cross sections are measured for charge-exchange, ionization, and
excitation within the same experimental setup for the Li$^{+}-$Ar, K$^{+}-$
Ar, and Na$^{+}-$ He collisions in the ion energy range $0.5-10$ keV.
Results of our measurements along with existing experimental data and the
schematic correlation diagrams are used to analyze and determine the
mechanisms for these processes. The experimental results show that the
charge-exchange processes are realized with high probabilities and electrons
are predominately captured in ground states. The cross section ratio for
charge exchange, ionization and excitation processes roughly attains the
value $10:2:1$, respectively. The contributions of various partial inelastic
channels to the total ionization cross sections are estimated and a primary
mechanism for the process is defined. The energy-loss spectrum, in addition,
is applied to estimate the relative contribution of different inelastic
channels and to determine the mechanisms for the ionization and for some
excitation processes of Ar resonance lines for the K$^{+}-$Ar collision
system. The excitation function for the helium, as well as for the sodium
doublet lines for the Na$^{+}-$He collision system, reveals some unexpected
features and a mechanism to explain this observation is suggested.
\end{abstract}

\pacs{34.80.Dp, 34.70.+e, 34.50.Fa, 32.80.Zb}
\maketitle

\section{INTRODUCTION}

The exploration of inelastic processes, with the goal of obtaining reliable
data on the corresponding cross section, are considered as a powerful tool
in understanding the dynamics taking place in slow ion$-$atom collisions.
Usually, models using a molecular basis set are applied to determine the
transition probabilities between the terms corresponding to the main
inelastic channels. However, such terms, due to the complicated many-channel
character, have been calculated for only a small number of simple collision
systems. Therefore, the schematic correlation diagrams of diabatic molecular
orbitals are widely used. However, they provide only a qualitative
explanation of the processes. The situation becomes more problematic from an
experimental point of view when particles with closed electron shells are
involved in the collision process. There are several reasons for this, but
mainly it is related to the following: 1) precise measurements of cross
sections for ionization and charge exchange processes require collection and
identification of all secondary particles; 2) a reliable determination and
control of the relative and absolute spectral sensitivity of the
light-recording system while performing an optical measurement is needed. It
is known that in such collisions these processes take place at relatively
small impact parameters corresponding, on avarage, to large momentum
transfers. Because of this the inelastic processes are realized when the
incident ions are scattered through relatively large angles and, due to
momentum conservation, this is accompanied by the formation of much more
energetic secondary particles than for weakly bound valence electrons. The
energy of these secondary particles (ions and free electrons) may reach tens
of electron volts, therefore, their full collection is problematic. At the
same time, these secondary particles along with primary particles are used
to determine cross sections. This circumstance has not been considered in
the earlier works \cite{Flaks1, Ogurtsov4, Mouzon9, Moe5} measuring cross
sections for ionization and charge-exchange processes. Therefore, some
doubts related to the reliability of these data and, hence, conclusion drawn
from them, seem to be indicated. For example, in case of ionization
processes, a standard capacitor method was used \cite{Flaks1} which yields
significant inaccuracies in measurements of cross sections, because no
measures were taken to prevent the scattering of primary and high-energy
secondary ions. As to charge-exchange processes, the magnitude of almost all
previously obtained cross sections \cite{Flaks1, Ogurtsov4, Moe5} are
underestimated, which is caused by the fact, that the method used for the
detection of neutral particles failed to ensure collection of those
particles which were scattered through large angles. Therefore, it is
necessary to reinvestigate some collision systems more accurately and some
for the first time. These studies require combined experimental approaches,
incorporating with the methods which are free of the deficiencies mentioned
above, and a variety of reliable experimental data for more persuasive
interpretation. From the other side one can use schematic correlation
diagrams to interpret the data and to identify the mechanisms leading to the
processes being investigated. Research on alkali-metal ion collisions with
rare gas atoms have been carried out by a variety of methods \cite{Flaks1,
Ogurtsov4, Mouzon9, Moe5, Matveev7, Matveev8, Afrosimov2, Francois13,
Lorents12, Latypov11, Kita18, V. V. Afrosimov16 21, Bidin6, Lomsadze2015,
Jrgensen20, Os22, Kita17, Kita2000, Lomsadze23, Kita10, KitaR, Kez25, Kita19}%
, however the available data for the absolute cross sections of the above
mentioned processes are not always consistent with each other \cite{Flaks1,
Mouzon9, Moe5, Latypov11, Bidin6} and in some cases, unreliable \cite%
{Ogurtsov4}.

The objectives of present work are detailed studies of processes induced by
an interaction of closed-shell alkali-metal ions of Li$^{+},$ K$^{+},$ and Na%
$^{+}$ with atoms of rare gases Ar and He. We performed measurements of
absolute differential and total cross sections for charge-exchange,
ionization and excitation processes in Li$^{+}-$Ar, K$^{+}-$Ar, and Na$^{+}-$%
He collisions in the range of ions energy 0.5$-$10 keV. In collisions of a Li%
$^{+},$ K$^{+},$ or Na$^{+}$ ion beam with Ar and He atoms we focus on the
following charge-exchange processes:

\begin{eqnarray}
\text{Li}^{\text{+}}+\text{Ar} &\rightarrow &\text{Li}+\text{Ar}^{+},
\label{Li} \\
\text{K}^{\text{+}}+\text{Ar} &\rightarrow &\text{K}+\text{Ar}^{+},
\label{K} \\
\text{Na}^{\text{+}}+\text{He} &\rightarrow &\text{Na}+\text{He}^{+},
\label{Na}
\end{eqnarray}%
where the products of the reaction can be in the ground states or in
different excited states. The collisions can induce the following ionization
processes:

\begin{eqnarray}
\text{Li}^{\text{+}}+\text{Ar} &\rightarrow &\text{Li}^{\text{+}}+\text{Ar}%
^{+}+e,  \label{Lie} \\
\text{K}^{\text{+}}+\text{Ar} &\rightarrow &\text{K}^{+}+\text{Ar}^{+}+e,
\label{Ke} \\
\text{Na}^{\text{+}}+\text{He} &\rightarrow &\text{Na}^{+}+\text{He}^{+}+e.
\label{Nae}
\end{eqnarray}%
\qquad \qquad The reactions (\ref{Lie}) - (\ref{Nae}) represent ionization
processes for the target atoms, that include different channels for the
excitation of the produced ions of the target atoms or/and incident ions, as
well as the excitation of autoionization states of the target atom that
leads to its ionization. In collisions of K$^{+}$ and Na$^{+}$ ions with Ar
and He atoms our study is focused on the following excitation processes: 
\begin{eqnarray}
\text{K}^{\text{+}}+\text{Ar} &\rightarrow &\text{K}^{\text{+ *}}+\text{Ar}%
^{\ast },  \label{K*} \\
\text{Na}^{\text{+}}+\text{He} &\rightarrow &\text{Na}^{\text{+ *}}+\text{He}%
^{\ast }.  \label{Na*}
\end{eqnarray}%
The excitation processes (\ref{K*}) and (\ref{Na*}) include different
channels for excitation of the K$^{+}$ or/and Ar atom, and the Na$^{+}$ ion
or/and He atom, respectively.

Charge-exchange cross sections for the K$^{+}-$He, K$^{+}-$Ne, K$^{+}-$Ar, K$%
^{+}-$Kr, K$^{+}-$Xe collision systems were reported in Ref. \cite{Ogurtsov4}
using the detection of fast neutral particles with a restricted interval of
scattering angles. However, as shown in Ref. \cite{V. V. Afrosimov16 21},
this limiting condition on the interval for collision angles in Ref. \cite%
{Ogurtsov4} underestimated the measured charge-exchange cross sections for
all collision systems and, particularly, for K$^{+}-$ Ar this
underestimation reaches more than a factor of ten. The absolute excitation
cross section of the potassium resonance spectral line for the K$^{+}-$Ar
collision is reported in Ref. \cite{Odom1977}, however it was measured just
for low energies, up to 0.7 keV. A convincing result, though in a limited
energy interval 1$-$3 keV, for the excitation cross section of summed
spectral Ar lines $\lambda $ $=104.8-106.7$ nm in K$^{+}-$Ar collision
system are presented in Ref. \cite{Bobashov1970}. Double differential cross
sections for the K$^{+}-$Ar collision system have also been measured in Ref. 
\cite{V. V. Afrosimov16 21}. An inelastic collision mechanism for the
systems of Li$^{+}-$He and Li$^{+}-$Ne has been studied experimentally and
theoretically in Ref. \cite{Francois13, Lorents12, Kita2000, Junker, Sidis}.
An absolute value of the differential cross section at two fixed energies, $%
E=200$ eV and $E=350$ eV of Na$^{+}$ ions, is reported in Ref. \cite{Kita18}%
. A relative differential cross section for the Na$^{+}-$Ar collision system
is measured in Ref. \cite{Kita17}. The most comprehensive study for
excitation mechanisms in Na$^{+}-$He and K$^{+}-$He collisions, at the ion
energies of $1.0-1.5$ keV and by the differential spectroscopy, were
reported recently in Ref. \cite{Kita19}. Double differential cross sections
were measured by detecting all scattering particles (Na$^{+}$, Na, K$^{+}$,
K, He$^{+}$, He) over a wide range of scattering angles.

For the Na$^{+}-$ Ar collision an energy spectrum of electrons ejected from
autoionization states of Ar atoms at $E=15$ keV \cite{Jrgensen20} have been
reported, but no result exist for low energy collisions. The result of the
measurements for the excitation function for the Na$^{+}-$He and K$^{+}-$He
collision systems, though in arbitrary units are reported in Refs. \cite%
{Matveev7, Matveev8}. A normalized emission cross section of He resonance
atomic lines for Na$^{+}-$He collisions are presented in Ref. \cite%
{Bobashev70}. Another comprehensive study of slow ion-atom collisions with
closed electron shells was carried out for the Na$^{+}-$Ne \cite{Os22},
although measurements were performed for a limited energy interval.

This brief survey of above mentioned processes occurring in alkali-metal ion
impact with atoms of rare gases shows that today there is no systematic
experimental measurements and reliable data available. Most of experimental
results are either provided in arbitrary units or reported for a restricted
energy interval. To our mind, the reason for this scarcity of measurements
is linked with the difficulties of performing the research with close
electron shell particles: scattering on large angles and problem of
collection of secondary particles. This circumstance motivates the present
detailed investigation of the primary mechanisms for these collision
processes. We have so far studied collisions between closed electron shell
particles for the K$^{+}-$He, Na$^{+}-$Ne, Na$^{+}-$Ar, Ne$^{+}-$Na, and Ar$%
^{+}-$Na collision systems \cite{Lomsadze2015, Lomsadze23, Kez25}. In this
paper, we report systematic studies of absolute total cross sections for
charge-exchange, ionization and excitation for the Li$^{+}-$Ar, K$^{+}-$Ar,
and Na$^{+}-$He collisions in a broad range of collision energies (0.5$-$10
keV), as well as the energy-loss spectra for the K$^{+}-$Ar collision
system. We have also measured the energies of the electrons liberated in the
collisions. We have found that for the K$^{+}-$Ar collision, the energy of
most of the liberated electrons are within the interval 20$-$32 eV, the
electrons' energy are less than 17 eV for the Na$^{+}-$He collision, and
below 12 eV for the Li$^{+}-$Ar collision system.

The remainder of this paper is organized as follows: in Sec. II. the
experimental techniques and measurement procedures are described and three
different experimental methods of measurements are presented. Here we
introduce the procedures for measuring of absolute total and differential
cross sections for charge-exchange, ionization and excitation processes. The
description of our measurements and the comparison of them with the results
of previously obtained experimental studies for the charge-exchange,
ionization and excitation occurring in the Li$^{+}-$Ar, K$^{+}-$Ar, and Na$%
^{+}-$He collisions are given in Sec. III. The discussion of experimental
results, and determination and clarification of the mechanisms for the
charge-exchange, ionization and excitation processes are presented in Sec.
IV. Finally, in Sec.V, we summarize our investigations and present the
conclusions.

\section{EXPERIMENTAL TECHNIQUES AND MEASUREMENT PROCEDURES}

\textit{2.1.} The main experimental set-ups used in the present experiments
for measurements of total and differential cross sections for
charge-exchange, ionization and excitation processes are the following: i.
the refined version of a capacitor method; ii. a collision spectroscopy
method; iii. an optical spectroscopy method. The basic approach for
measurements of inelastic processes realized in collision of an alkali-metal
ion with rare gases atoms was described previously in Ref. \cite{Kez25}, so
only the details of the apparatus with description outline of methods and
measurement procedures will be given here. A beam of Li$^{+}$ , K$^{+}$, or
Na$^{+}$ ions from a surface-ionization ion source is accelerated, formed,
and focused by an ion-optics system, which includes quadruple lenses and
collimation slits \cite{Kez25}. After the beam passes through a magnetic
mass spectrometer, it enters the collision chamber containing target Ar or
He gasses. The pressure in the collision chamber is kept at about 10$^{-6}$
Torr, while the typical pressure under operation with the Ar and He target
gasses is a two order of magnitude less. This ensures single-collision
conditions for the ion-target atom collision. The charge-exchange and
ionization cross section were measured by a refined version of the capacitor
method \cite{Kikiani26}. This method allows preventing the electrodes by the
scattering primary ions that affect the results of measurement. The
secondary positive ions and free electrons produced during the collision are
collected and detected by a collector. The collector consists of two rows of
plate electrodes that ran parallel to the primary ion beam. A uniform
transverse electric field, that extracts and collects secondary particles,
is created by the potential applied to the grids. This method represents a
direct measurement of the yield of produced singly positively charged ions
and free electrons as the primary beam passes through the target gas.
Obviously, the measured quantities are related to the capture cross section
and the apparent ionization cross section \cite{Kez25}. The uncertainties in
the measurements of the charge-exchange cross sections for the Li$^{+}-$Ar, K%
$^{+}-$Ar, and Na$^{+}-$ He collision systems are estimated to be 7\%, 15\%,
and 10\%, respectively, while the uncertainties in the measurements of the
ionization cross sections are estimated to be 12\%, 10\%, and 15\%,
respectively. These uncertainties are determined primarily by the
uncertainties in the measurements of absolute value of the cross sections
for production of positively charged ions and free electrons, as the primary
beam passes through the target, and by the uncertainty in the measurement of
the target gas pressure in the collision chamber.

\textit{2.2.} The energy-loss spectra and differential scattering
experiments are performed by the collision spectroscopy method. Since the
details of the apparatus have been given elsewhere \cite{Gochitashvili27},
only a brief description is presented here. The primary beam extracted from
the ion source was accelerated to the desired energy before being analyzed
according to $q/m$ ($q$ and $m$ are the ion's charge and mass,
respectively). The analyzed ion beam after being collimated by a slit enters
into the collision chamber and then passes into a \textquotedblleft
box\textquotedblright\ type electrostatic analyzer. The energy resolution of
this analyzer is $\Delta E/E$ $=1/500$. The voltage applied to the analyzer
is scanned automatically which allows to investigate the energy-loss spectra
in the energy range of $0-100$ eV. The differential cross section is
measured by rotating the analyzer around the center of collision over an
angular range between 0$^{0}$ and 20$^{0}$. The laboratory angle is
determined with respect to the primary ion beam axis with an accuracy of 0.2$%
^{0}$. Such a tool gives us the option to determine the total cross sections
and compare them with the results obtained by the refined version of the
capacitor method \cite{Kikiani26}. In addition, the measured energy-loss
spectra provide detailed information related to the intensity of inelastic
processes realized in the charge-exchange, ionization and excitation
reactions.

\textit{2.3.} The method used for the optical measurements was discussed
previously in Ref. \cite{Gochitashvili28}, therefore, only brief description
will be given here. The alkali-metal ion beam leaving the surface-ionization
ion source after acceleration to a predetermined energy is focused by the
quadruple lenses and analyzed by a mass-spectrometer. The emerging ions
passes through a differentially pumped collision chamber with the target gas
at low pressure. The light emitted from the collision chamber as a result of
the excitation of colliding particles, is viewed perpendicularly to the beam
by an optical spectrometer. The spectral analysis of the radiation is
performed in the vacuum ultraviolet and in the visible spectral regions. The
linear polarization of the emission in the visible part of the spectra is
analyzed by a Polaroid and a mica quarter-wave phase plate. A
photomultiplier tube with a cooled cathode is used to analyze and detect the
emitted light. The spectroscopic analysis of the emission in the vacuum
ultraviolet spectral region is performed with the Seya -- Namioka vacuum
monochromator incorporating a toroidal diffracting grating. The radiation is
recorded by the secondary electron multiplier used under integrating or
pulse-counting conditions. The polarization of the radiation in the vacuum
ultraviolet region is not taken into account. We determine the absolute
excitation cross sections by comparing the measured output signal with the
one that is obtained due to the excitation of nitrogen molecule by electron
impact. Particular attention is, therefore, devoted to the reliable
determination and control of the relative and absolute spectral sensitivity
of the light-recording system in the visible part of spectra. The latter is
done by measuring the photomultiplier output signal due to the bend of the
first negative system of the ion N$_{2}^{+}$ (B$^{2}\Sigma u^{+}-$X$%
^{2}\Sigma g^{+}$ transition) and due to the Mainel system (A$^{2}\Pi u^{+}-$%
X$^{2}\Sigma g^{+}$ transition) excited in collisions between the electrons
and nitrogen molecules \cite{Kezer2010}. These bands cover the wavelength
interval between $423.6$ nm and $785.4$ nm. The output signal is normalized
to the $(0.1)$ band ($\lambda =427.8$ nm) which has the highest intensity in
this range. The relative spectral sensitivity of the light recording system
obtained in this way is compared with the measured relative excitation cross
sections for the same bands, averaged over the experimental data reported in
Refs. \cite{Ajello29, Avakyan30, Stone31, Tan33, Pendelton1972,
Srivastava1968}. The absolute uncertainties in the excitation cross section
for the K$^{+}-$Ar and Na$^{+}-$He colliding system are estimated to be 15\%
and the uncertainty of relative measurements is about 5\%.

\section{RESULTS OF EXPERIMENTAL MEASUREMENTS}

\textit{3.1.} The measured energy dependences of the absolute cross section
for the charge-exchange, ionization and excitation processes are presented
in Figs. $1-3$. Figure 4 represents a typical example of the energy-loss
spectrum for K$^{+}-$Ar collisions. Figs. $1-3$ reveal marked differences
between the energy dependences of the cross sections for the various
processes. While the ionization cross section shown in Fig. 2 increases
monotonically with energy, the charge-exchange cross section has almost a
flat energy dependence, and the excitation function exhibits an oscillatory
structure. Another feature is the magnitude of the cross sections. Among of
the processes investigated the largest value of cross section has
charge-exchange processes. The data for the charge-exchange cross sections
for the Li$^{+}-$Ar, K$^{+}-$Ar, and Na$^{+}-$He collisions with electron
capture to the ground state of potassium K$(4s)$ state for the K$^{+}-$Ar,
lithium Li$(2s)$ state for the Li$^{+}-$ Ar, and sodium Na$(3s)$ state for
the Na$^{+}-$He collision systems, respectively, are presented in Fig. 1. In
the same figure are presented the charge-exchange in the excited resonance K$%
(4p)$ state, for the K$^{+}-$Ar, and Na$(3p)$ state, for the Na$^{+}-$He.
For comparison we also present in Fig. 1 the data of other authors. The
comparison of our charge-exchange cross sections for the K$^{+}-$Ar
collision sytem (curve 1a) with the results from Ref. \cite{KitaPhysB} at a
fixed impact energy $E=350$ eV (open square) and results obtained in Ref.%
\cite{V. V. Afrosimov16 21} at a fixed impact energy $E=2$ keV (open
triangle) shows satisfactory agreement within the experimental
uncertainties. However, a dramatic difference, by about two orders of
magnitude as well as in the shape of the energy dependences are observed
when one compares our results for the K$^{+}-$Ar system (curve 1a) with data
from Ref. \cite{Flaks1} (curve 1b). The same tendency, but with a smaller
discrepancy of one order of magnitude, is observed when one compares our
results for the Li$^{+}-$Ar (curve 2a) and Na$^{+}-$He with the results from
Ref. \cite{Ogurtsov4} (curves 2b and 3b, respectively). Our results for the
Na$^{+}-$He charge-exchange processes (curve 3a) can be compared with the
cross section obtained in differential scattering experiments \cite{Kita17}
at energy of $E=1.5$ keV, (open circle). This comparison shows that here the
discrepancy only amounts to a factor of three. At and below $1.5$ keV impact
energy our excitation cross section for K$(4p)$ state in the K$^{+}-$ Ar
collision (curve 1c) in energy region $E<1.5$ keV is in satisfactory
agreement with the earlier data \cite{Odom1977} (curve 1d).

\begin{figure}[tbp]
\label{label__figure_1} \includegraphics[width=10cm,angle=0]{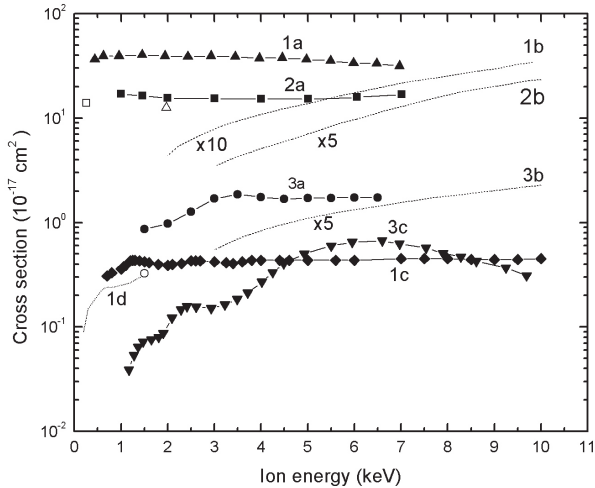} 
\vspace{0 cm}
\caption{Dependences of the absolute charge-exchange cross sections on
energy of Li$^{+}$, K$^{+}$ and Na$^{+}$ ions in Li$^{+}-$Ar , K$^{+}-$Ar,
and Na$^{+}-$He collisions. Results of the present study in comparison with
the previous measurements from Refs. \protect\cite{Flaks1, Ogurtsov4, V. V.
Afrosimov16 21, Kita19, Odom1977, KitaPhysB}. Curves: 1a $-$ electron
capture in the ground potassium K$(4s)$ state for the K$^{+}-$ Ar system,
present data; 1b $-$ electron capture in ground potassium K$(4s)$ state for
the K$^{+}-$Ar, data from Ref. \protect\cite{Flaks1} are multiplied by a
factor of 10; 1c $-$ electron capture in the excited resonance K$(4p)$ state
($\protect\lambda =766.5-769.8$ nm, transition $4p-4s$) for the K$^{+}-$Ar
collision, present data ; 1d $-$ electron capture in excited resonance K$(4p)
$ state $(\protect\lambda =766.5$ nm, transition $4p-4s)$ for the K$^{+}-$%
Ar, data from Ref. \protect\cite{Odom1977}; $\square $ $-$ electron capture
in the ground potassium K$(4s)$ state, for the K$^{+}-$Ar system, data from
Ref. \protect\cite{KitaPhysB}; $\triangle $ $-$ electron capture in the
ground potassium K$(4s)$ state for the K$^{+}-$Ar, data from Ref. 
\protect\cite{V. V. Afrosimov16 21}; 2a $-$ electron capture in ground
lithium Li$(2s)$ state for the Li$^{+}-$Ar system, present data; 2b $-$
electron capture in ground lithium Li$(2s)$ state for the Li$^{+}-$Ar sytem,
data from Ref. \protect\cite{Ogurtsov4} are multiplied by a factor of 5; 3a $%
-$ electron capture in ground sodium Na$(3s)$ state for the Na$^{+}-$He
system, present data; 3b $-$ electron capture in ground sodium Na$(3s)$
state for the Na$^{+}-$He, data from Ref. \protect\cite{Ogurtsov4} are
multiplied by a factor of 5; $\bigcirc $ $-$ electron capture in ground
sodium Na$(3s)$ state for the Na$^{+}-$He system, data from Ref. 
\protect\cite{Kita19}; 3c $-$ electron capture in excited resonance sodium
Na $(3p)$ state $(\protect\lambda =589.0-589.6$ nm, transition $3p-3s)$ for
the Na$^{+}-$He, present data.}
\end{figure}

\textit{3.2.} The data for the ionization cross section for the Li$^{+}-$Ar,
K$^{+}-$Ar, and Na$^{+}-$He collisions, along with the previously obtained
measurements, are shown in Fig. 2. Here curve 1a represents ionization cross
section for the K$^{+}-$Ar, curve 2a -- for the Li$^{+}-$Ar and curve 3a --
for the Na$^{+}-$He collision systems. The comparison of our ionization
cross section for the K$^{+}-$Ar collision (curve 1a) with the results
obtained in Ref. \cite{Flaks1} (curve 1b) and the results obtained at a
fixed $E=2$ keV impact energy in Ref. \cite{Afrosimov2} (open triangle), as
well as comparison of our results for the Li$^{+}-$Ar collision system
(curve 2a) with the results from Ref. \cite{Flaks1} (curve 2b) shows a
satisfactory agreement in the energy range studied. Satisfactory agreement
are observed also at low energies while comparing our results for the Na$%
^{+}-$He (curve 3a) with the result from Ref. \cite{Flaks1} (curve 3b), but
the discrepancy increases for ion energies $E>5$ keV. A rather significant
discrepancy is observed if one compares our results for K$^{+}-$Ar (curve
1a) with the previously obtained results in Ref. \cite{Moe5} (curve 1c). Our
experimental results for the ionization cross section for the Li$^{+}-$Ar
collision system (curve 2a) are in an excellent agreement with theoretical
prediction \cite{Solov'ev40} (curve 2c).

\begin{figure}[tbp]
\label{label__figure_2} \includegraphics[width=10cm,angle=0]{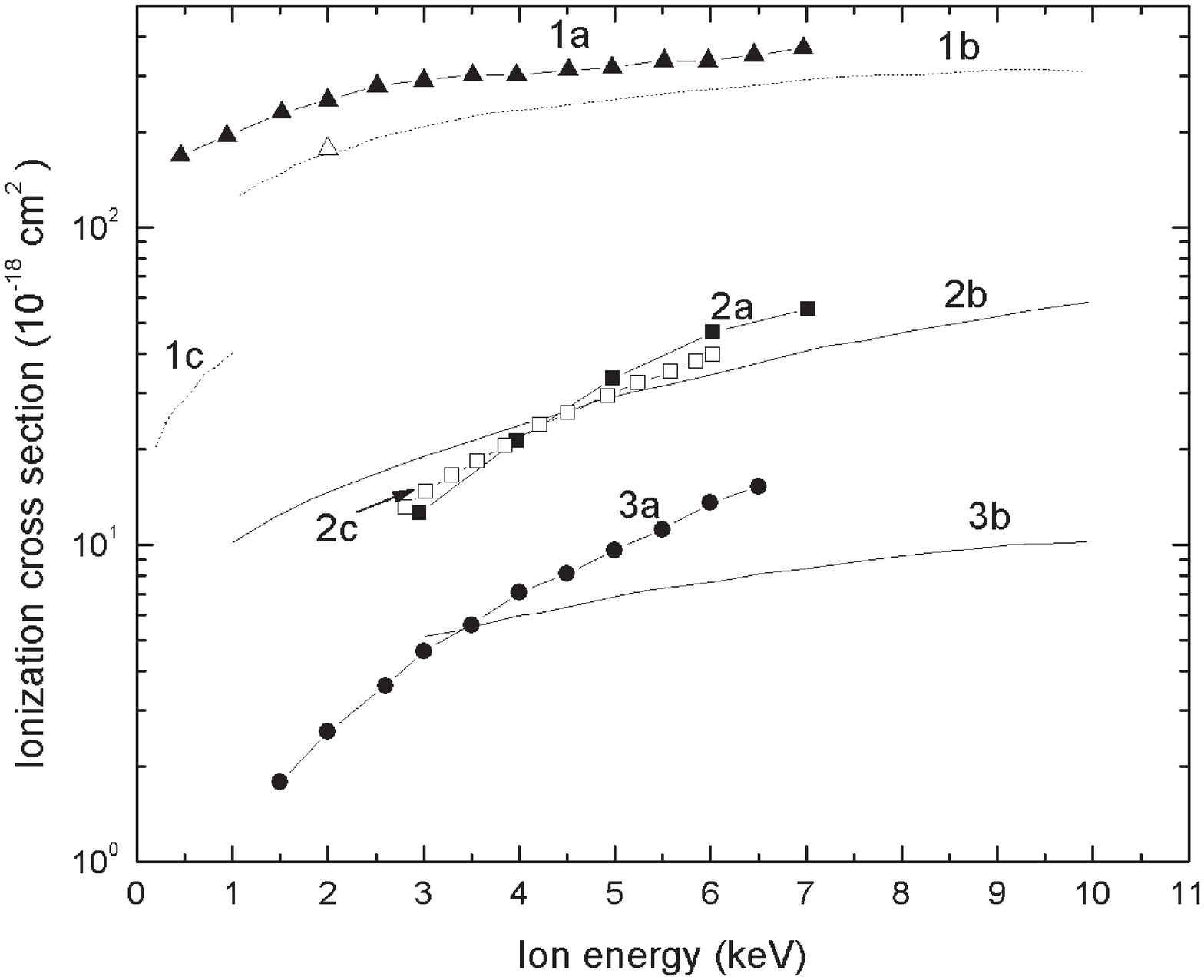} 
\vspace{0 cm}
\caption{Dependences of the absolute ionization cross sections on energy of
Li$^{+}$, K$^{+}$, and Na$^{+}$ ions in the Li$^{+}-$Ar, K$^{+}-$Ar, and Na$%
^{+}-$He collisions. Results of the present study in comparison with the
previous measurements from Refs. \protect\cite{Flaks1, Moe5, Afrosimov2,
Solov'ev40}. Curves: 1a -- K$^{+}-$Ar, present data; 1b - K$^{+}-$Ar, data
from Ref. \protect\cite{Flaks1}; 1c- K$^{+}-$Ar, data from Ref. \protect\cite%
{Moe5}; $\triangle $- K$^{+}-$Ar, data from Ref. \protect\cite{Afrosimov2};
2a -- Li$^{+}-$Ar, present data; 2b -- Li$^{+}-$Ar, data from Ref.%
\protect\cite{Flaks1}; 2c- Li$^{+}-$Ar, data from Ref. \protect\cite%
{Solov'ev40}; 3a -- Na$^{+}-$He, present data; 3b - Na$^{+}-$He, data from
Ref. \protect\cite{Flaks1}.}
\end{figure}

\textit{3.3.} The results of measured excitation cross sections realized for
the K$^{+}-$Ar and Na$^{+}-$He collisions, along with the data obtained by
other authors, are presented in Fig. 3. The data obtained in our study for
the excitation function of the argon atomic resonance line ($\lambda =106.7$
nm, $3p^{5}4s-3p^{6}$ transition) emitted in the K$^{+}-$Ar collision are
presented by curve 1a. The excitation cross section for the resonance
spectral Na doublet lines, $\lambda =589.0$ nm and $\lambda =589.6$ nm, $%
3p-3s$ transition, in the Na$^{+}-$He collision is presented by curve 3a,
while the resonance helium atomic line, $\lambda =58.4$ nm, $2p-1s$
transition, and the sum of the excitation cross section of sodium and helium
lines in collisions between sodium ions and helium atoms, are presented by
curves 3a, 2a, and 4a, respectively.

\begin{figure}[tbp]
\label{label_figure_3} \includegraphics[width=10cm,angle=0]{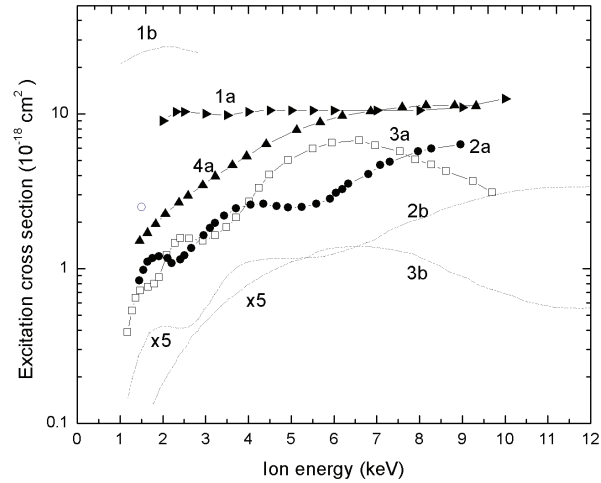} 
\vspace{0 cm}
\caption{Excitation cross sections for the resonance argon atomic line for
the K$^{+}-$Ar collision system, and for the resonance lines of sodium
doublet and helium atom in the Na$^{+}-$He collision. Results of the present
study in comparison with the previous measurements from Refs. \protect\cite%
{Kita19, Bobashov1970, Bobashev70, Maurer1938}. Curves: 1a - excitation
function of the resonance argon atomic line $\protect\lambda =106.7$ \ nm , $%
3p^{5}4s-3p^{6}$ corresponding to the transition $3p^{5}4s-3p^{6}$ in the K$%
^{+}-$Ar collision system, present data; 1b - excitation function of the
summed resonance argon atomic lines $\protect\lambda =104.8$ nm and $106.7$
nm \ corresponding to the transions $3p^{5}4s-3p^{6}$ and $%
3p^{5}4s^{^{\prime }}-3p^{6}$ for the K$^{+}-$Ar collision sytem, data from
Ref. \protect\cite{Bobashov1970}; 2a, 2b and $\bigcirc $- excitation
function of the resonance helium atomic line $\protect\lambda =58.4$ nm
corresponding to the $2p-1s$ transition in the Na$^{+}-$He collision system,
present data, data from Ref. \protect\cite{Bobashev70} multiplied by a
factor of 5 and data from Ref. \protect\cite{Kita19}, respectively; 3a and
3b - excitation function of the resonance sodium doublet lines $\protect%
\lambda =589.0$ nm and $589.6$ nm for the transition $3p-3s$ in the Na$^{+}-$%
He system, present data and data from Ref. \protect\cite{Maurer1938}
multiplied by a factor 5, respectively; 4a -- the summed excitation cross
sections of the resonance helium atomic line (curve 2a) and resonance sodium
doublet lines (curve 2b), for the Na$^{+}-$He collision, present data.}
\end{figure}

Comparison of our data for the excitation function for the sodium doublet
and for the helium atom with the results reported in Refs. \cite{Maurer1938,
Bobashev70} shows, that there is a considerable discrepancy in magnitude of
the excitation cross sections. Moreover, significant discrepancies are
observed also in the energy dependence of the excitation cross section for
the sodium doublet with the exception of the position of the maximum at $%
E=6.5$ keV. Our observations show an oscillating structure of the cross
section which was not seen in Ref. \cite{Maurer1938}. Our results for the
energy dependence with a pronounced oscillatory structure on the excitation
cross section of the helium atom are the same as in Ref. \cite{Kita2000} and
the oscillations are in phase, but there are discrepancies in the magnitude.
The discrepancies, as compared with Ref. \cite{Maurer1938}, are due to the
considerable experimental uncertainties introduced by the photographic
method employed in Ref. \cite{Maurer1938} to record the radiation. This is
also the explanation of the discrepancy between the magnitudes of our
results and the results obtained in Ref. \cite{Maurer1938}, (curve 3b) since
the data from Ref. \cite{Maurer1938} were used in Ref. \cite{Bobashev70} to
determine the absolute value of these cross sections. The comparison of our
results for the excitation of the Ar resonance $4s,$ line $\lambda =106.7$
nm (curve 1a) with the results from Ref. \cite{Bobashov1970}, (curve 1b) is
difficult because the data in Ref. \cite{Bobashov1970} are obtained in the
limited energy interval from 1 to 3 keV and they considered excitation of
the Ar resonance $4s$ as the sum of lines $\lambda =104.8$ nm and $\lambda =$
$106.7$ nm. The difference between our results of the helium atomic 
resonance line (curve 2a) and those from Ref. \cite{Kita19} obtained at
fixed energy, $E=1.5$ keV (open circle) reaches threefold.

\textit{3.4. }We studied the energy-loss spectrum for the K$^{+}-$Ar
collision system and a typical example of the inelastic energy-loss spectrum
for K$^{+}-$Ar collisions is presented in Fig. 4. The same spectrum was
presented in our previous paper \cite{Kez25} for reference. In the present
study this spectrum will be used to investigate which mechanism is dominant
in the K$^{+}-$Ar collisions. Generally, in our study for this collision
system, we measured spectra through different fixed angles in the range $%
1^{0}-7^{0}$. However, the spectrum in Fig. 4 is chosen for a fixed energy
of $E=2$ keV and scattering angle $\theta =$3.5$^{0}$ at which the inelastic
channels are well pronounced. It is seen from figure that the spectrum has a
discrete character. The first peak of this spectrum with zero energy-loss
corresponds to the elastically scattered ions. The second peak in the
spectrum corresponds to the single electron excitation of the argon atom
into $3p^{5}4s$, $3p^{5}4p$ and $3p^{5}3d$ states with the energy-loss $Q$
within $11.6-14.0$ eV, and a single ionization of the Ar atom in the state $%
3p^{5}$at the energy-loss of $15.7$ eV. The third peak correspond to the
excitation of K$^{+}$ ion into the $4s$ and $3d$ states with the energy-loss 
$16<Q<22$ eV, while the fourth one to the excitation of autoionization
states of the Ar atom with the excitation of one $3s$ electron or two $3p$
electrons, and to the ionization with the excitation of the Ar, with the
energy-loss of $25<Q<32$ eV. An investigation of the dependence of the area
of each of the peaks versus the scattering angle of the incident ions has
shown that the excitation cross sections of the investigated transitions
exhibits an onset behavior. Acritical s the scattering angle is increased,
the cross section of each inelastic transition remains small, up to a
certain scattering angle $\theta _{c}$. When $\theta _{c}$ is reached the
cross section of the transition increases sharply. With further increase of
the scattering angle the cross sections of the inelastic transitions
decreases. Such onset behavior of the angular dependences of differential
cross sections for the inelastic transitions makes it possible to integrate
these cross sections for the estimation of relative contribution of
different inelastic channels, and compare them with the obtained total cross
section. In our study such approach has been applied to determine the
mechanism for the ionization processes and for some excitation processes of
Ar atoms realized in the K$^{+}-$Ar collision that is presented in Sec. IV.

\begin{figure}[tbp]
\label{label_figure_4} \includegraphics[width=8cm,angle=0]{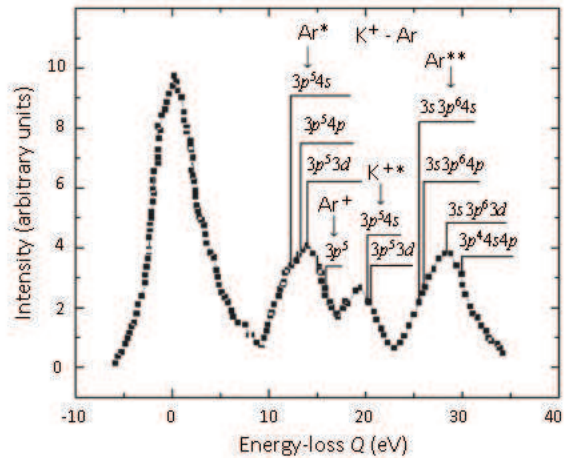} \vspace{%
0 cm}
\caption{Typical energy-loss spectrum in laboratory reference system for K$%
^{+}-$Ar collisions measured at $E=2$ keV collision energy and $\protect%
\theta =3.5^{0}$ scattering angle.}
\end{figure}

\section{DISCUSSION OF EXPERIMENTAL RESULTS AND MECHANISMS OF PROCESSES}

\subsection{Charge exchange in Li$^{+}-$Ar , K$^{+}-$Ar, and Na$^{+}-$He
collision systems}

We use the experimental data for the individual inelastic channels along
with the corresponding schematic correlation diagrams to establish the
mechanism for the charge-exchange processes for the Li$^{+}-$Ar, K$^{+}-$Ar,
and Na$^{+}-$He collision systems.

\subsubsection{Li$^{+}-$Ar collisions}

The results of the charge-exchange cross sections for the Li$^{+}-$Ar
collision system are presented in Fig. 1 by curve 2a. It is seen that the
dependence of the cross sections on the energy of Li$^{+}$ ions is weak. No
oscillations are seen in the cross section, and its average value amounts to
1.4$\cdot $10$^{-16}$ cm$^{2}$.

The inelastic differential cross section measurements for the Li$^{+}$ ion
collisions with rare gases have been performed in the keV energy range in
Ref. \cite{Barat73}. Besides, in that work using a single electron
configuration approximation, the calculations are reported for the potential
curves of $\Sigma $ symmetry, corresponding to the ground state of the
system and to the one and two electron excitation states of colliding
particles. From these calculations it follows that there are two coupling
regions in which non-adiabatic transitions may cause charge-exchange: one
occurs at internuclear distance $R\thicksim $1.5 a.u., and the other at $%
R\leq $0.5 a.u. Also the authors suggested, that in both cases the most
probable mechanism responsible for the charge-exchange process is the
transition from the ground state of the system to the state corresponding to
the electron capture into the ground state of lithium Li($1s^{2}2s)$ as it
is seen from the correlation diagram in Fig 5. The transition for the
non-adiabatic region, corresponding to the internuclear distance $R\thicksim
1.5$ a.u., has been also observed in experiments \cite{Barat73} by measuring
differential cross sections. Unfortunately, the transition region at $R\leq
0.5$ a.u. was not investigated in Ref. \cite{Barat73}, since the scattering
was studied up to 25 keV$\cdot $deg., which means that the particles
approach each other only up to $R\geq 0.75$ a.u. For estimation of the
contribution of the transition at $R\leq 0.5$ a.u. to the total
charge-exchange cross section, we integrate the differential cross section
given in Ref. \cite{Barat73} at energy of $E=3$ keV and compare with our
result. The comparison shows that the results coincide well each other
within the experimental uncertainties. Therefore, one can conclude that the
contributions from transitions corresponding to $R\leq 0.5$ a.u. to the
charge-exchange cross section is negligible and the contribution of the
transition at $R\thicksim 1.5$ a.u. to the charge-exchange cross section is
the dominant one. A possible reason for the importance of this region, where
the terms of X$^{1}\Sigma $ and A$^{1}\Sigma $ energetically approach each
other, is the prevalence of the attraction between the Ar$^{+}$ ion and Li
atom due to polarization over the repulsion caused by the exchange
interaction at those distances.

\subsubsection{K$^{+}-$Ar collisions}

To determine the mechanism for the K$^{+}-$Ar collisions we compare the
total charge-exchange cross section presented in Fig. 1 by curve 1a with
those corresponding to the decay of resonance levels of the potassium atom,
presented by curve 1c. Taking into account the selection rules and the ratio
of oscillator strengths for the transitions, we show that the decay of any
of the excitation levels of the potassium atom culminates in about half the
cases with a transition of the atom to a resonant state, which then decays.
Thus, the doubled de-excitation cross section of the potassium atom gives a
clue related to the capture cross section in the excited state. The
comparison of this two sets of our results show that the contribution to the
cross section from capture of an electron to the excited $4p$ state of
potassium atom to the total charge-exchange cross section is small, so that
the main contribution is provided by the capture to the $4s$ ground state.
However, the energy dependence of these cross sections is the same: the
cross sections reach their maxima at low K$^{+}$ energies $E\sim 1$ keV and
vary slowly in a wide energy range. The experimental results for the K$^{+}-$%
Ar collision that leads to the charge-exchange can be explained
qualitatively in terms of the schematic correlation diagrams for molecular
orbitals given in Ref. \cite{Barat}. The analysis of this diagram shows that
the capture to the ground $4s$ and excited $4p$ states of the potassium atom
can occur as a result of a transition between terms of the same $\Sigma
-\Sigma $ symmetry. The processes responsible for the population of these
states are competing processes. This conclusion is supported by the fact
that the terms corresponding to these states are populated from the same
initial state term. Moreover, the parameters of the quasi-crossing region
are such that the emission maxima occur for equal velocities. Substantial
discrepancies between the magnitudes of the cross section are linked to the
crossing point $R$ of molecular terms and can probably be explained by the
fact that the processes are occurred in different crossing points $R$ of
molecular terms. The initial state term at first populates the term
corresponding to the ground $4s$ state at large internuclear distance $R$,
and after that to the excited $4p$ state at comparatively smaller distances $%
R$. These considerations lead to the conclusion that charge-exchange into
the ground $4s$ state and excited $4p$ state of the potassium atom involves
the Landau-Zener type interaction and occurring in a quasi-crossing region
between initial K$^{+}(3p^{6})-$Ar$(3p^{6})$ and charge transferred K$(4s)$
+ Ar$(3p^{6})$ and/or K$(4p)$ + Ar$(3p^{6})$ terms of $\Sigma -\Sigma $
symmetry.

\begin{figure}[tbp]
\label{label_figure_5} \includegraphics[width=10
cm,angle=0]{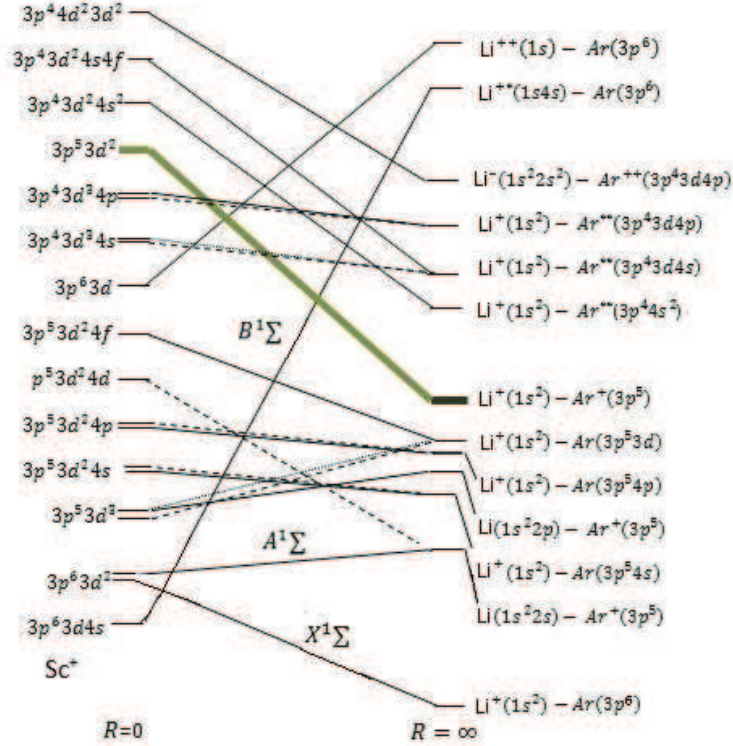} \vspace{0 cm}
\caption{Schematic correlation diagram for some states of the Li$^{+}-$Ar
collision system. Solid lines indicate $\Sigma $ states, dashed lines
indicate $\Pi $ states and dotted lines indicate $\Delta $ states.}
\end{figure}

\subsubsection{Na$^{+}-$He collisions}

The same procedure, as for the K$^{+}-$Ar collision is applied to determine
the mechanism for charge exchange processes occurring in Na$^{+}-$He
collisions. Two sets of experimental results are obtained in our study: one
by the refined version of the capacitor method for measuring the total
charge exchange cross sections of the prosesses

\begin{eqnarray}
\text{Na}^{\text{+}}+\text{He} &\rightarrow &\text{Na}(2p^{6}3s)+\text{He}%
^{+}(1s),  \label{Ne1} \\
&\rightarrow &\text{Na}(2p^{6}3p)+\text{He}^{+}(1s),  \label{Ne2} \\
&\rightarrow &\text{Na}(2p^{6}3d)+\text{He}^{+}(1s),  \label{Ne3}
\end{eqnarray}%
when the sodium atom is in various states shown by curve 3a and the second
one by optical measurements for the excitation into the states of Na($%
2p^{6}3p$)$+$He$^{+}+21.5$ eV (curve 3c). The comparison of these results
(curve 3a and curve 3c in Fig. 1) shows that the fraction of the total cross
section related to the sodium excited atom via reaction (\ref{Ne2}) amounts
roughly to 10\% of the total cross section. Therefore, the charge-exchange
processes in this study is mostly attributed to the reaction (\ref{Ne1})
with an electron capture to the ground state of the projectile. In Ref. \cite%
{Kita19} an estimate was performed, though for the fixed energy $E=1.5$ keV,
of the revealing contribution of various inelastic channels realized in the
Na$^{+}-$He collisions to charge-exchange processes. The authors of Ref. 
\cite{Kita19} show that the direct one electron target excitation in the
reaction Na$^{+}-$He$^{\ast }(1s2s)+20.6$ eV and charge exchange into the
ground 3s state of Na$(3s)$ + He$^{+}+19.44$ eV occur with high
probabilities. The charge-exchange processes to the ground $3s$ state of
sodium atom can occur, as one can see from the correlation diagram in Fig.
6, due to the direct pseudo-crossing of the term corresponding to the state
Na$(2p^{6}3s)+$He$^{+}(1s)$ with the initial state of the system Na$%
^{+}(2p^{6})+$He$(1s^{2})$. Since the Na$^{+}(2p^{6})+$He$(1s^{2})$ state
has only $\Sigma $ symmetry it follows that the $\Sigma -\Sigma $ transition
play a dominant role in the charge-exchange processes.

\begin{figure}[tbp]
\label{label_figure_6} \vspace{0.0 cm} %
\includegraphics[width=10cm,angle=0]{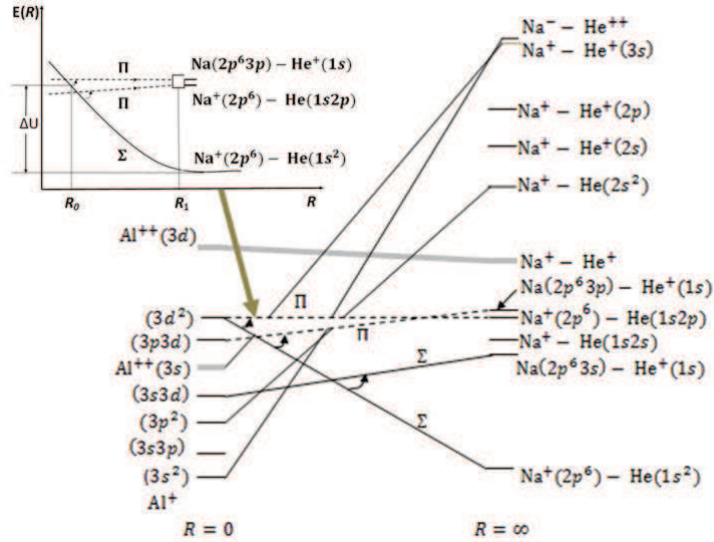}
\caption{Schematic correlation diagram for some states for the Na$^{+}-$He
collision system. Solid lines indicate $\Sigma $ states; dashed lines
indicate $\Pi $ states. The insert at the upper left corner shows the
quasi-crossing of the ground state term of the system Na$^{+}(2p^{6})-$He$%
(1s^{2})$ with the Na$^{+}(2p^{6})-$He$(1s2p)$ and Na$(2p^{6}3p)-$He$%
^{+}(1s) $ terms. The arrow indicates the corresponding terms at the
schematic correlation diagram at small internuclear distances..}
\end{figure}

\subsection{Ionization in Li$^{+}-$Ar, K$^{+}-$Ar, and Na$^{+}-$He collisions%
}

\subsubsection{Li$^{+}-$Ar collisions}

To discuss the mechanism responsible for the ionization in Li$^{+}-$Ar
collisions, we construct a schematic correlation diagram shown in Fig. 5
according to the rules presented in Ref. \cite{Barat}, electron energy
spectra obtained in our study and the data of differential cross section
presented in Ref. \cite{Barat73}. This diagram is used for identification of
the ionization mechanism. It follows from the results of the present
measurements of the energy of the electrons ejected in Li$^{+}-$Ar
collisions, that the liberation of mostly slow electrons with energies less
than 10 eV is a characteristic of the ionization in Li$^{+}-$Ar collisions.
In order to determine the mechanism responsible for this feature, we
estimated the contribution of several inelastic processes which result in
the emission of slow electrons to the total ionization cross section. In
order to estimate the fraction coming from the autionization states of Ar
atoms to the ionization cross section, the data for differential cross
section obtained in Ref. \cite{Barat73} have been used. Integration of
differential cross section shows, that the contribution of these processes
to the total ionization cross sections amounts to just about 17\%. Small
contributions from atomic autoionization to the total ionization processes
follow also from analyses of the correlation diagram in Fig. 5. As it is
seen that none of the autoionization terms have an immediate crossing point
with the system of ground states and hence, as one can expect, their
population occurs with small probabilities.

To the best of our knowledge there are no experimental data in the
literature which could be used to make estimates of other contributions to
total ionization cross section. In order to estimate the contribution of
direct ionization Li$^{+}(1s^{2})+$Ar$(3p^{6})\rightarrow $Li$^{+}(1s^{2})+$%
Ar$^{+}(3p^{5})+e$, which in the quasimolecular model is linked to the
transition of an adiabatic term into the continuum, we use the procedure
described in Ref. \ \cite{Solov'ev40}. Analysis of the correlation diagram
of the molecular orbital for the (LiAr)$^{+}$ system in Fig. 5 shows that in
the united atom limit, the $3p$ state of the Ar atom evolves into the $3d$
state of the Sc$^{+}$ ion. Since this result is of importance for evaluation
of the contribution of the direct ionization, we note that it agrees with
the Barat-Lichten correlation rules \cite{Solov'ev40}, and the Eichler-Wille
rules \cite{Eichler1976}. Consequently, the value $l=2$ was chosen for the
evaluation of the cross section. The binding energy $E_{nl}$ of the
electrons in the nonadiabatic region was chosen to be equal to that of the $%
3d$ electrons of the Sc$^{+}$ ion. The effective charge $Z_{eff}$ was
determined by the interpolation of the data from Ref. \cite{Hartree41}, and
for the $3d$ electrons of the Sc$^{+}$ ion $Z_{eff}=$4.8. The estimate of
the cross section for the direct ionization with these parameters are in
reasonable agreement with our measurements (curves 2a and 2c in Fig. 2): the
absolute value of the cross sections are closer and the energy dependences
are similar.

The same procedure was used to estimate the contribution of stripping
processes, that are also the source of electrons, to the ionization cross
section. The correlation diagram in Fig. 5 indicates that the $1s$ state of
the Li$^{+}$ ion correlates with the $3d$ state of the scandium ion Sc$^{+}$
in the united atom limit. The binding energy of these electrons is known 
\cite{Bearden67} and hence future calculation become possible. Estimates of
the contribution for the stripping processes with these parameters show,
that at the ion energy of 3.0 keV the contribution of the stripping
processes to the total ionization cross section is less than 0.4\%, while at
6.0 keV it is less than 2\%. Therefore, this contributions of the stripping
processes are insignificant over the entire energy range studied.

Although the above mentioned agreement of the estimation for the direct
ionization with the experiment is fortuitous to some extent, this result
together with the estimate of portion for the atomic autoionization and the
stripping allows us to conclude, that the contribution of direct ionization
process to the total cross section for the electron emission is the
governing mechanism for the Li$^{+}-$Ar collisions.

\subsubsection{K$^{+}-$Ar collisions}

It follows from the results of the present measurements of the electron
ejection cross sections that the liberation of electron with energies in the
range $20<E_{e}<32$ eV strongly contribute to the ionization in K$^{+}-$Ar
collisions. As the result for energy-loss spectrum for the K$^{+}-$Ar
collision (see Fig. 4) among the different processes that lead to ionization
of the target atom with above mentioned energy of electrons, a decisive role
may be played by the processes of autoionization and of ionization with
excitation of the Ar atom. From the energy-loss spectrum plotted in Fig. 4
it is seen that these processes are represented by the fourth peak of that
spectrum. As it will be shown below, contribution of direct ionization
(direct removal of an electron to the continuum spectrum with formation of
an intermediate autoionization state) and multiple ionization of the target
atom, make no noticeable contribution to the effective cross section for the
target atom ionization. In order to determine the channel and mechanism of
ionization we estimate the contribution of direct ionization K$^{+}(3p^{6})+$%
Ar$(3p^{6})\rightarrow $K$^{+}(3p^{6})+$Ar$^{+}(3p^{5})+e$ that in the
quasimolecular model is linked to the transition of the adiabatic term to
the continuum. We estimate the cross section of this process using the
results obtained in Ref \cite{Solov'ev40}. It should be noted that the
expression given in Ref. \cite{Solov'ev40} is written in terms of the
principal quantum number for a hydrogen-like ion. Therefore, we modify it
slightly for the estimation of the cross section for the emission of
electrons from multi-electron atoms. The final expression which is used here
can be found in our previous work \cite{Kez25}. We use this expression for
the interpretation of the results for the Li$^{+}-$Ar and it will be
employed for the Na$^{+}-$He collision systems in the next subsection.

An analysis of correlations for molecular orbitals in one-electron
approximation for the K$^{+}-$Ar system shows that the $3p$ electrons of the
Ar atom, whose ionization is considered, in the limit of the united atom
correspond to the $4d$ electrons of Rb$^{+}$ ion. Thus, for the estimation
we chose the orbital angular momentum $l=2$. The binding energy $E_{nl}$ of
electrons, in the nonadiabaticity region, was assumed to be equal to the
binding energy of $4d$ electrons of the Rb$^{+}$ ion. Since it would be
difficult to determine accurately the effective charge $Z_{eff}$ and binding
energy $E_{nl}$ for the $4d$ electrons of the Rb$^{+}$ ion, being in the
excited states with the electron configuration of Ar like ion [Ar]$%
3d^{6}4p^{2}4f^{2}4d^{5}5f^{2}$, the effective charge $Z_{eff}$ was
determined by the interpolation of the results\emph{\ }obtained by a Hartree
method \cite{Hartree41} in Ref. \cite{Barat73}. Estimates of the cross
section for the direct ionization with these parameters show, that at an ion
energy of $E=2$ keV the contribution of the direct ionization to the total
ionization cross section is less than 1.5\%, while at $E=7.0$ keV it is less
than 2.5\% and, hence, its average value does not exceed 3.5$\cdot $10$^{-18}
$ cm$^{2}$. This means that this contribution is insignificant over the
entire energy range studied. According to these calculations and based on
the results, obtained in our study for the energy of liberated electrons, we
can conclude that the primary mechanism for the ionization is connected with
the decay of autoionization states. As seen from the energy-loss spectrum
for the K$^{+}-$Ar collision system (Fig. 4), indeed the states with
energies of $25-34$ eV are excited with noticeable probability. These
energies correspond to the autoionization states of Ar atom with excitation
of one $3s$ electron and two $3p$ electrons with configurations of $%
3s3p^{6}4s$ and $3p^{4}$($^{1}$D)$4s$($^{2}$D)$4p(^{1}$P$)$, respectively.

\subsubsection{Na$^{+}-$He collisions}

The liberation of electrons with the energies less than 17 eV is a
characteristic of ionization processes in Na$^{+}-$He collisions as it
follows from the results of present measurements of emitted electrons. In
order to reveal the channels and establish the mechanism responsible for the
ionization, we estimate the contribution of several inelastic processes that
result in emission of slow electrons. Using the Barat -- Lichten rules \cite%
{Barat} and the data from Ref. \cite{Kita19} we construct the correlation
diagram presented in Fig. 6. The contribution of the direct ionization Na$%
^{+}(2p^{6})$ + He(1s$^{2}$)$\rightarrow $Na$^{+}(2p^{6})$ + He$^{+}(1s)+e$
is estimated following Ref. \cite{Solov'ev40}. An analysis of the
correlation diagram of molecular orbitals for the Na$^{+}-$He system shows,
that the $1s$ state of the He atom, whose ionization is considered, in the
limit of the united atom becomes the $2p$ state of the Al$^{+}$ ion. Thus,
for the estimate of cross section we chose the orbital angular momentum $l=1$%
. The binding energy $E_{nl}$ of the electrons in the nonadiabatic region
was chosen to be equal to the $2p$ electrons of the Al$^{+}$ ion (28.75 eV).
The effective charge $Z_{eff}$ was determined by the interpolation of the
data obtained by a Hartree calculation\emph{\ }in Ref. \cite{Hartree41}. For
the $2p$ electrons of the Al$^{+}$ ion, we obtain $Z_{eff}=3.7$. The
estimate of the cross section for the direct ionization with these
parameters shows, that at an ion energy of $1.5$ keV the contribution of the
direct ionization to the total ionization cross section is less than 5.5\%,
while at $6.5$ keV it is less than 7\%.

The same procedure is applied to determine the contribution of the electron
yield to the measured ionization cross section as a result of stripping of
the projectile ions. As it follows from the diagram in Fig. 6 the $2p$ state
of the Na+ ion is correlated with the $3d$ state of the Al$^{+}$ ion.
Consequently, for calculation of the cross section we select $l=2$, while we
take $Z_{eff}$ to be the same as for the $2p$ electrons of the Al$^{+}$ ion
used in evaluating the ionization cross section. As a result of this
calculation we find that the contribution of stripping to the total electron
emission cross section is less than 12\% for the energy of 1.5 keV of the
sodium ions and does not exceed 17\% for the energy of 6.5 keV.
Consequently, we can conclude that the contribution of the stripping
processes as well as direct ionization processes to the total ionization
cross section is insignificant in the entire energy range interval studied.
Other possible sources for liberation of the electrons like e.g. double
ionization of the He atom, Na$^{+}+$He$\rightarrow $Na$^{+}+$He$^{2+}+2e$,
and direct two-particle excitation Na$^{+}+$He$\rightarrow $Na$^{+^{\ast
}}(2p^{5}3s)+$He$^{\ast }(1s2s$) and capture accompanied by the ionization
of the He$^{+}$ ion that is also a possible source for electron liberation
via reaction Na$^{+}+$He$\rightarrow $Na$+$He$^{+^{\ast }}\rightarrow $Na$+$%
He$^{2+}+e,$ evidently make a small contribution to the ionization cross
section. There are two reasons for this. The absence of pseudo-crossings of
the corresponding quasimolecular terms with the ground state term, as it
seen from the diagram in Fig. 6, and the large energy defect for these
processes, 54 eV, 53.9 eV and 48.9eV respectively. Consequently, by
systematically evaluating the contribution of various inelastic processes to
the ionization of the target atoms in Na$^{+}-$He collisions, we find that
the ionization may be caused primarily by the decay of autoionization states
in an isolated atom. This assessment is supported by the correlation diagram
in Fig. 6 and by the data obtained in Ref. \cite{Kita19} as well. According
to Ref. \cite{Kita19} these states are those with two excited electrons
associated with a direct two electron excitation of the He atom in
autoionization state: Na$^{+}+$He$\rightarrow $Na$^{+}+$He$^{\ast \ast
}(2s^{2})\rightarrow $Na$^{+}+$He$^{+}+e$.

\subsection{Excitation processes in K$^{+}-$Ar and Na$^{+}-$He collisions}

\subsubsection{K$^{+}-$Ar collisions}

In discussing the mechanism for the Ar$(3p^{5}4s)$ excitation processes in K$%
^{+}-$Ar collisions we use the energy-loss spectrum obtained in our study
and results reported in Ref. \cite{Afrosimov2}. In Ref. \cite{Afrosimov2}
the area corresponding to the energy-loss range of $11.5-16$ eV has been
integrated and for the sum of Ar excited states $3p^{5}4s$, $3p^{5}4p$ and $%
3p^{5}3d$ we obtained a value of 7$\cdot $10$^{-17}$ cm$^{2}$. Since the
above mentioned excited levels are energetically close as seen in Fig. 4,
the determination of the relative probabilities of their excitation
encounters considerable difficulties. Moreover, as shown by a qualitative
analysis of the energy-loss spectrum (the shape of the peak and the position
of its maximum), the levels for the configurations of $3p^{5}4p$ and $%
3p^{5}3d$ are excited with high probability, while the level for the
configuration $3p^{5}4s$ is populated with a surprisingly small probability
although the level of $3p^{5}4s$ state lays at lower energy than the levels $%
3p^{5}4p$ and $3p^{5}3d$, ($11.5-12$ eV and $13-14$ eV respectively, see
Fig. 4). This fact and also the requirement to have reliable data related to
the individual channel, that includes the clarification of the contribution
of various excitation processes and determination of mechanisms for the
processes, motivate us to perform optical measurements for the resonance Ar$%
(3p^{5}4s)$ line in the reaction (\ref{K*}).

The analysis of the spectrum for the K$^{+}-$Ar collision system presented
in Fig. 4 shows that the second peak in the energy-loss spectrum with the
width of $11.5-16$ eV corresponds not only to the excitation processes, but
also to the direct ionization of the Ar atom (the ionization potential $15.7$
eV). Our estimate shows that, the value of direct ionization cross section
for the K$^{+}-$Ar collision system amounts to a magnitude of 2.8$\cdot $10$%
^{-18}$ cm$^{2}$. As seen from our results given by curve 1a in Fig. 3 the
average excitation cross section for the resonance Ar line in the $3p^{5}4s$
state amounts to 1$\cdot $10$^{-17}$ cm$^{2}$. Thus, due to the smallness of
the direct ionization cross section as well as the excitation cross section
for the resonance Ar$(3p^{5}4s)$ line, compared to the sum of the excitation
cross sections 7$\cdot $10$^{-17}$ cm$^{2}$, one can conclude that the
dominant role in the Ar excitation processes is played by the excitation of
the $3p^{5}4p$ and $3p^{5}3d$ states. As for the mechanism for the
excitation of resonance $3p^{5}4s$ state of the Ar atom, its population is
caused by a cascade transition from already mentioned upper laying levels $%
3p^{5}4p$ and $3p^{5}3d$ to the resonance $4s$ level.

\subsubsection{Na$^{+}-$He collisions}

The oscillatory structure of the energy dependence of the cross sections is
most pronounced for the Na$^{+}-$He collision system. Our results show that
the oscillation on the excitation cross sections for the resonance lines of
sodium and helium (curve 3a and curve 2a in Fig. 3) atoms are in antiphase.
The curve obtained by adding of the excitation cross sections of these lines
turns out to be smooth over the entire energy range (curve 4a in Fig. 3).
This means that the observed oscillations are a consequence of interference
of two energetically neighboring vacant excited states (the difference
between the states is 0.03 eV) of the systems Na$^{+}(2p^{6})+$He$(1s2p)$
and Na$(2p^{6}3p)+$He$^{+}(1s)$ interacting at large internuclear distance.

On the other hand, for the realization of a quasi-molecular interference
phenomenon it is necessary that both of these excited terms to be populated
by the same entrance term. As seen from the correlation diagram in Fig. 6
this term may correspond to the ground state Na$^{+}(2p^{6})+$He$(1s^{2})$,
which in the limit of united atom promotes to the autoionization state of
aluminum ion Al$^{+}(3d^{2})$. Therefore, the ground state term, which
possess $\Sigma $ symmetry, populates $\Pi $ terms of the Na$(2p^{6}3p)$ + He%
$^{+}(1s)$ and Na$^{+}(2p^{6})$ +He$(1s2p)$ states at small internuclear
distance due to the rotational $\Sigma -\Pi $ transition and then at a large 
$R$ an interaction of the terms with the same $\Pi -\Pi $ symmetry take
place.

The experimental results for the excitation function for the resonance lines
of sodium and helium atoms, as well as the mechanisms defined for these
processes, allows us to obtain additional information regarding the
restoration of parameters for the interaction area. For this reason we use
the procedure suggested in Ref. \cite{Tolk1973}. The insert in Fig. 6 shows
the quasi-crossing of the ground state term of the system Na$^{+}(2p^{6})-$He%
$(1s^{2})$ with the Na$^{+}(2p^{6})-$He$(1s2p)$ and Na$(2p^{6}3p)-$He$%
^{+}(1s)$ terms at small internuclear distances. Taking into consideration
that a quasi-crossing of ground state term and the later excitation terms
that are populated at small internuclear distance \cite{Bobashev78}, the
experimental data allow to determine the mean value of threshold energy $%
\Delta U$ for the excited states at the average internuclear distance $R_{0}$
and the area of \textquotedblleft rectangular loop\textquotedblright\ formed
by the terms of these states, shown in the insert in Fig. 6. The area of
\textquotedblleft rectangular loop\textquotedblright , which consists of the
terms of these states, is $\left\langle \Delta E\cdot R\right\rangle $,
where $\Delta E$ is the splitting energy of the terms and $R$ is the
internuclear distance. The estimate shows that $\Delta U$ =70 eV and the
value of $\left\langle \Delta E\cdot R\right\rangle \thicksim 1.4\cdot $10$%
^{-7}$ eV$\cdot $cm. The value of $\Delta U$ corresponds to the excitation
energy at the quasi-crossing of the ground state term and terms of
interference states. Using a potential energy curve for the ground state of
the system (NaHe)$^{+}$ from Ref. \cite{Nikulin1975} and the value $\Delta U=
$70 eV determined from our data, one can unambiguously obtain that the
quasi-crossing of the terms occurs at $R\thicksim $0.5 $\mathring{A}$
internuclear distance.

It is well established that for a small internuclear distance $\Sigma -\Pi $
transitions, which are related to a rotation of the internuclear axis, are
the most important. Under such condition, the absolute value of the cross
section should be small and by increasing the velocity of relative motion of
particles it should increase, reaching maxima at comparatively large
velocities. Our data confirm this assumption. Moreover, comparatively large
oscillation depth, observed in our study, and the data reported in Ref. \cite%
{Bobashev76} support the assumption that at large internuclear distances the
interaction of the terms with the same $\Pi -\Pi $ symmetry take place,
which itself are populated due to the rotational transition.

\section{SUMMARY AND CONCLUSIONS}

In this study we report the measurements for absolute differential and total
cross sections for charge-exchange, ionization and excitation processes in Li%
$^{+}-$Ar, K$^{+}-$Ar, and Na$^{+}-$He collisions in the energy range of 0.5$%
-$10 keV. We have also measured the energies of the electrons liberated in
the collisions and found that the energy of most of the liberated electrons
are below 12 eV for the Li$^{+}-$Ar collision, within the interval $20-32$
eV for the K$^{+}-$Ar collision system, and less than 17 eV for the Na$^{+}-$%
He collision. The measurements are performed under the same experimental
conditions, using a refined version of the condenser plate method, the
collision spectroscopy method with angle- and energy-dependent detection of
the collision products, and the optical spectroscopy method, with an
accurate calibration procedure of the light recording system. The comparison
of our measusrements with existing experimental and theoretical results are
presented.

\bigskip We construct the correlation diagrams for the (LiAr)$^{+}$ and
(NaHe)$^{+}$ systems based on the rules formulated in Ref. \cite{Barat}. The
experimental data and the schematic correlation diagrams are used to analyze
and determine the mechanisms for the charge-exchange, ionization and
excitation processes for the reactions (\ref{Li})-(\ref{Na*}). We found that
the charge-exchange processes occur with high probabilities and electrons
predominately are captured to the ground states of the resultant atom in the
region of pseudocrossing of potential curves of $\Sigma $ symmetry. The
contribution to the charge-exchange cross section for the Li$^{+}-$Ar
collision from the transition corresponding to $R\thicksim 1.5$ a.u. is the
dominant one. The results of experimental studies show that the cross
sections ratio for charge-exchange, ionization and excitation processes
roughly attains the value $10:2:1$, respectively. The contributions of
various partial inelastic channels to the total ionization cross section are
estimated and primary mechanism for this process is defined. The
contribution of the direct ionization process to the total cross section for
the electron emission is the governing mechanism for the Li$^{+}-$Ar
collision. The primary mechanism for the ionization in the K$^{+}-$Ar
collision is connected with the decay of autoionization state. Our results
confirm the conclusion of Ref. \cite{Kita19} that \ the ionozation in the Na$%
^{+}-$He collision is related to a direct two electron excitation of the He
atom in autoionization state: Na$^{+}+$He$\rightarrow $Na$^{+}+$He$^{\ast
\ast }(2s^{2})\rightarrow $Na$^{+}+$He$^{+}+e$.

The energy-loss spectrum is applied to estimate the relative contribution of
different inelastic channels and to determine the mechanisms for the
ionization and for some excitation processes. The main mechanism for the
excitation of resonance $3p^{5}4s$ state of the Ar atom in the K$^{+}-$Ar
collision is related to the cascade transition from upper laying levels $%
3p^{5}4p$ and $3p^{5}3d$ to the resonance $4s$ level. The oscillatory
structure of the energy dependence of the cross sections is most pronounced
for the Na$^{+}-$He collision system. This behavior is observed in the
excitation function for the helium resonance line ($\lambda =$ 58.4 nm, $%
2p-1s$ transition) and for the sodium doublet lines $\lambda =$ $589.0$ nm
and $\lambda =$ $589.6$ nm, $3p-3s$ transition in Na$^{+}-$ He collisions.
We conclude that this phenomenon is a consequence of interference of two
energetically separated by 0.03 eV excited states of the Na$^{+}(2p^{6})$ +
He$(1s^{2})\rightarrow $Na$(3p)+$ He$^{+}(1s)$ and Na$^{+}(2p^{6})+$He$%
(1s^{2})\rightarrow $Na$^{+}(3p^{6})+$ He$(1s2p)$ systems at large
internuclear distances.

\section*{Acknowledgements}

This work was supported by the Georgian National Science Foundation under
the Grant No.31/29 (Reference No. Fr/219/6-195/12). R.Ya.K. and R.A.L
gratefully acknowledge support from the International Reserch Travel Award
Program of the American Physical Society, USA.

\end{document}